\documentstyle[psfig]{laa}

\def\kms{\relax \ifmmode {\,\rm km\,s}^{-1}\else \,km\,s$^{-1}$\fi}
\def\ha{\relax \ifmmode {\rm H}\alpha\else H$\alpha$\fi}
\def\hb{\relax \ifmmode {\rm H}\beta\else H$\beta$\fi}

\def\lha{\relax \ifmmode L_{{\rm H}\alpha}\else $L_{{\rm H}\alpha}$\fi}
\def\shi{\relax \ifmmode \sigma_{{\rm HI}}\else $\sigma_{\rm HI}$\fi}
\def\sh2{\relax \ifmmode \sigma_{{\rm H}_2}\else $\sigma_{{\rm H}_2}$\fi}

\def\Msun{M_\odot}

\def\ref#1{{\hangindent=\parindent \hangafter=1 \par \noindent #1}}
\def\degr{\hbox{$^\circ$}}
\def\arcmin{\hbox{$^\prime$}}
\def\arcsec{\hbox{$^{\prime\prime}$}}
\def\deg{\hbox{$^\circ$}}
\def\min{\hbox{$^\prime$}}

\def\fdg{\hbox{$.\!\!^\circ$}}
\def\fs{\hbox{$.\!\!^{\rm s}$}}
\def\farcm{\hbox{$.\mkern-4mu^\prime$}}
\def\farcs{\hbox{$.\!\!^{\prime\prime}$}}
\def\degd#1.#2{ #1\fdg#2 }                 
\def\mind#1.#2{ #1\farcm#2 }               
  
\def\kms{\,kms$^{-1}$} \def\mhz{\,MHz} \def\mpc{\,Mpc} 
\def\pc{\,pc} \def\mjy{\,mJy} \def\kel{\,K} 
\def\secd#1.#2{ #1\farcs#2 }               
\def\hhh{\ifmmode {\rm ^h}              
         \else {${\rm ^h}$} 
         \fi}
\def\sss{\ifmmode {\rm ^s}              
         \else {${\rm ^s}$}
         \fi}

\def\hms#1h#2m#3s{                      
                  \relax
                  \ifmmode #1^{\rm h}\,#2^{\rm m}\,#3^{\rm s}
                  \else \hbox{$#1^{\rm h}\,#2^{\rm m}\,#3^{\rm s}$}
                  \fi
                 }
\def\dms#1d#2m#3s{                      
                  \relax
                  #1\degr\,#2\arcmin\,#3\arcsec
                 }
\def\hmsd#1h#2m#3.#4s{                  
                      \relax
                      \ifmmode #1^{\rm h}\,#2^{\rm m}\,#3\fs#4
                      \else \hbox{$#1^{\rm h}\,#2^{\rm m}\,#3\fs#4$}
                      \fi
                     }
\def\apj#1, {{\it ApJ,~}{\bf#1}, }
\def\apjlett#1, {{\it ApJL,~}{\bf#1}, }
\def\apjsupp#1,{{\it ApJS,~}{\bf#1}, }
\def\aj#1, {{\it AJ,~}{\bf#1}, }
\def\astrf#1, {{\it Astrofizika,~}{\bf#1}, }
\def\aasupp#1, {{\it AAS,~}{\bf#1}, }
\def\aa#1, {{\it AA,~}{\bf#1}, }

\def\mnras#1, {{\it MNRAS,~}{\bf#1}, }
\def\mmnras#1, {{\it MemRAS,~}{\bf#1}, }
\def\annrev#1, {{\it ARAA,~}{\bf#1}, }
\def\ass#1, {{\it Astrophys.\ Space\ Sci.~}{\bf#1}, }
\def\pasp#1, {{\it PASP,~}{\bf#1}, }

\begin{document}
  \title{ Molecular gas in the barred spiral M~100}
  \subtitle{II. $^{12}$CO(1--0) interferometer 
observations and numerical simulations}


  \author{S. Garc\'\i a--Burillo$^{1}$, M.J.Sempere$^{1}$, F. Combes$^{2}$ 
and R. Neri$^{3}$}
%

\offprints{S. Garc\'\i a--Burillo}
\institute{ Observatorio Astron\'omico Nacional, Campus Universitario, 
Apdo 1143, E--28800 Alcal\'a de Henares (Madrid), Spain \and Observatoire de Paris, 
DEMIRM, 61 Av. de l'Observatoire, F--75014, France \and IRAM, 300 Rue de 
la Piscine, F-38406, France}
  \date{October 1997}
%
%
  \thesaurus{ 07.14.1, 07.02.1, 07.21.1, 07.11.1,09.19.1}
\maketitle

\begin{abstract}

Using the IRAM interferometer we have mapped at high resolution ($2\farcs2 \times 1\farcs2$) 
the $^{12}$CO(1--0) emission  in the nucleus of the doubled barred SABbc spiral M~100. Our synthesized map
includes the zero spacing flux of the single--dish 30m map (Sempere \& Garc\'\i a--Burillo, 1997, {\bf paper I}).
Molecular gas is distributed in a two spiral arm structure starting from the end points of the nuclear bar ($r=600$ pc)
up to $r=1.2$ kpc,  and a  central source ($r\sim$100 pc).
The kinematics of the gas indicates the existence of a steep rotation curve (v$_{rot}$=180 km\, s$^{-1}$
at $r\sim 100$ pc) and strong streaming motions characteristic of a trailing spiral wave inside corotation.

Interpretation of the CO observations and their relation with stellar and gaseous 
tracers (K, optical, H$\alpha$, H\,I and radiocontinuum maps) are made in the light of a numerical model
of the clouds hydrodynamics. Gas flow simulations analyse the gas response to a 
gravitational potential derived from the K-band plate, including the two nested bars. We develop two families
of models: first, a single pattern speed solution shared by the outer bar+spiral and by the nuclear bar,
 and secondly, a two independent bars solution, where the nuclear bar is dynamically decoupled and 
rotates faster than the primary bar. 

We found the best fit solution consisting of a fast pattern ($\Omega_f$=160 kms$^{-1}$kpc$^{-1}$) for the 
nuclear bar (with corotation at R$^{F}_{COR}$=1.2 kpc) decoupled from the slow pattern 
of the outer bar+spiral ($\Omega_f$=23 kms$^{-1}$kpc$^{-1}$) (with 
corotation at R$^{S}_{COR}$=8-9 kpc). As required by non-linear coupling 
of spirals (Tagger et al 1987), the corotation of the fast pattern falls in
the ILR region of the slow pattern,
allowing an efficient transfer of molecular gas towards the nuclear region.
Solutions based on a single pattern hypothesis for the whole disk 
cannot fit the observed molecular gas response and fail to account for the 
relation between other stellar and gaseous tracers.
In the two-bar solution,  the gas morphology and kinematics are strongly
varying in the rotating frame of the slow large-scale bar, and fit the
data periodically during a short fraction (about 20\%) of the relative nuclear 
bar period of 46 Myr.

\keywords{Galaxies: kinematics and dynamics of - Galaxies: barred -
Galaxies: spiral - Galaxies: individual - Interstellar medium: molecules}

\end{abstract}
\section{Introduction}

The advent of high-sensitivity near-infrared imaging of galaxies has shown that 
a significant percentage of barred spirals host secondary bars in their nuclei.
 There could be two interpretations of the {\it bars within bars} phenomenon,
according to the relative pattern speeds of the two bars (Friedli and 
Martinet, 1993; Friedli and Benz, 1993 and 1995; Combes, 1994). The two patterns
could be corotating if they are about parallel or perpendicular to 
each other. If the secondary inner bar is strongly misaligned with the primary 
outer bar, the two bars are likely to have distinct wave pattern speeds,
as shown by numerical simulations. The decoupling of an inner faster pattern 
appears in self-consistent simulations with gas and stars thanks to the role 
of the dissipative component: as a result of gas inflow, under the action of 
the bar gravitational torques, mass accumulates onto the 
x$_2$ types of orbits which weakens the primary bar. The rotation period
becomes much shorter in the nuclear regions due to mass concentration, which 
leads to the decoupling of a fast-rotating bar.
 Eventually, the nuclear bar destroys itself or it destroys the primary bar, 
modifying the overall disk potential. Evolution can then occur in much less 
than a Hubble time, and galaxies change their morphological type along
the Hubble sequence. They change from barred to un-barred, and also they
concentrate mass in the process, evolving slowly from late-types to
early types.

The observation and modelling  of {\it real} barred galaxies offers the 
opportunity to test theory predictions on galaxy evolution and it seems a 
necessary complement to numerical simulations of {\it model} galaxies. 
The present work is intended to bring a combined observational and modelling
 effort in the study of the nearby barred spiral M100 (NGC4321). In this 
galaxy, classified as SABbc by de Vaucouleurs et al (1991), the hypothesis of a 
single mode common to the whole disk is dubious, both from observational and 
theoretical evidences. 

\bigskip

On the observational side, 
the nuclear region of M100 (up to r=3kpc) has been so far the subject of 
numerous studies. The pioneering work of Arsenault and 
collaborators (1988, 1989, 1990) established 
a connection between the ring-like H$\alpha$ morphology of the nucleus and the 
existence of Inner Linblad Resonances.
Further steps in sensitivity made appear, first, a four-armed structure 
(Cepa et al, 1990) and recently a fragmented two spiral arm 
structure in H$\alpha$ (Knapen et al, 1996). The 6cm radio-continuum VLA 
map of Weiler et al (1981) shows also a two arm spiral pattern.    

Near infrared images of the nucleus (Pierce 1986, Shaw et al 1995, Knapen 
et al 1995 (hereafter {\bf K95}), Rauscher 1995)
show the existence of either a secondary nuclear bar in K (nearly parallel to
 the main bar) together with a leading spiral structure, or an inner oval in 
the I band (with principal axes misaligned with
respect to the outer bar).     
The synthesis aperture $^{12}$CO(1--0) maps (Canzian, 1992; Rand, 1995; 
Sakamoto et al 1995) indicate the existence of a two-arm molecular spiral structure 
connected to the K nuclear bar end points. The IRAM 30m map of {\bf paper I} 
shows a strong concentration of CO emission towards
the nuclear disk {\bf ND}, a component clearly distinguishable 
from the main bar. A steep rotation curve gradient, unresolved by 
the 30m beam (12\arcsec\, in the 2--1 line), indicates a high mass 
concentration in the {\bf ND}.  
 
\bigskip

On the modelling  side, 
Garc\'\i a-Burillo et al (1994) (hereafter  called {\bf GB94}) and Sempere 
et al (1995) (hereafter {\bf S95}) made numerical 
simulations of the cloud hydrodynamics to study the evolution of the 
molecular gas disk under the action of a realistic spiral+barred potential derived
from a red band plate. The authors assume the whole disk to be fitted by a single 
well defined wave pattern characterized by $\Omega_p$, shared by the primary bar and 
the spiral arms. However they lacked
first, of the necessary spatial resolution and secondly, of a fair potential tracer to 
analyse the gas response in the inner 500 pc.

{\bf K95} have made numerical simulations of the stellar and 
gas dynamics in M100, using a {\it model} potential which departs 
markedly from the real mass distribution. 
Although they favour a one bar mode scenario 
their model fails to reproduce the molecular gas distribution observed by the 
interferometer.   

\bigskip

 We present here a combined single-dish and interferometer data set 
fulfilling both high resolution (2.2\arcsec$\times$1.2\arcsec) and sensitivity 
requirements. Contrary to the synthesis aperture maps so far published, we 
recover entirely the zero-spacing flux of the {\bf ND}. The comparison between 
the different gaseous and stellar tracers of the {\bf ND} is reexamined in 
this work. Particular attention is paid to the bias introduced by extinction in 
optical and even near-infrared images of the nucleus, and what might be the 
implications on the interpretation of the data. 	
Observations are confronted to the result of new numerical simulations of the 
clouds hydrodynamics, based on a mass distribution 
directly derived from the infrared luminosity image of M100.
 We will focus on the feasibility of two independent patterns in the disk and 
how this scenario accounts better 
for the observations.

\section{Observations}

$^{12}$CO(1--0) visibilities were obtained between
February 1995 and April 1996 by the IRAM Plateau de Bure interferometer.  
We used the 4-antennas set of extended configurations
(BC) supplemented by 3-antennas observations compact set (D).  
All antennas were equipped with SIS receivers operating in the single-side-band 
mode and yielding receiver temperatures close to
40\kel. The system temperatures range typically from 250\kel\, to
400\kel. The cross-correlator was adjusted within a
540\mhz\, passband with a channel width of 2.5\mhz, corresponding to a
velocity resolution of 6.5\kms\, in the $^{12}$CO(1--0) line.

Instrumental calibration was performed using the quasars 
3C273 and 1219+285. The atmospheric rms phase
fluctuations on the longest baselines, integrated over 4 minutes, were
always found to be $\le 50$\,degrees. The receiver passband was
verified on 3C273. The absolute flux density scale was
based on measurements of 3C273 and 1219+285. Information on their fluxes were
bootstrapped from regular IRAM monitoring observations (Dutrey \&
Ungerechts, 1995, 1996). The data were antenna-based calibrated using 
the CLIC package (Lucas, 1992).

Short spacing visibility data was provided by the 30m
radiotelescope. Single-dish visibilities were derived from the
brightness temperature distribution following a reduction scheme described in 
the software package GILDAS (Guilloteau \& Forveille, 1989) (see also the
comprehensive discussions in Vogel et al.\ 1984, Neri et al.\ 1997). Firstly,
the single-dish data were resampled to the 2.5\mhz\, frequency
resolution of the interferometric data, and the main beam
temperatures were converted to Jy/beam by applying the standard
conversion factor $S/T_A^* = 5.5$\,Jy/K adopted for the 30m telescope
(Wild 1995). Secondly, the single-dish maps were deconvolved by
dividing their Fourier transforms by the Fourier transform of the
21$''$ beam of the telescope, which was assumed to be circular and
Gaussian, and the results multiplied by the 42$''$ primary beam of the
interferometer. Finally, the single-dish map was corrected from a 2$\arcsec$
absolute residual pointing error. The derived M100 dynamical center (see below) coincides 
with the emission peak in the single-dish and the interferometer 
synthesized field.

Single-dish and interferometric visibilities were then combined 
into a common visibility table by scaling the single-dish weights
to the mean weight of the short spacing interferometer visibilities in
the {\em uv} range $15-25m$. Henceforth we implicitly correct for 
eventual calibration errors in the 30m data.

CLEANed maps were then obtained from the visibilities by standard
deconvolution procedures (no tapering and uniform weighting). The map field
is $256 \times 256$ pixel$^{2}$ in size with a linear scale of 
$0\farcs25$/pixel. The synthesized beam was $2\farcs2 \times 1\farcs2$ and 
oriented along PA=30\,degrees. The rms noise level in the 2.5\mhz
channel maps is $\sim4$mJy/beam or equivalently 140mK within the
synthesized beam. Point source sensitivity, determined for the
line-free 400\mhz continuum passband is $\sim 0.65$\mjy/beam.

We adopted $\alpha$(1950)= 12$^h$20$^m$23.2$^s$, $\delta$(1950)=
16\deg 06\arcmin 00\arcsec
as the phase tracking center. Correlator was centered 
at v$_{LSR}$=1562\kms. However, we re-determined the dynamical center
and the systemic velocity of NGC4321, following a standard least
squares procedure applied to the interferometer data cube. Assuming a priori a
position angle of PA=153\deg ({\bf paper I}), the dynamical center (the (0,0) offset
as referred in this work) is found at $\alpha$(1950)=12$^h$20$^m$22.96$^s$,
$\delta$(1950)=16\deg 05\arcmin 57.5\arcsec ; the systemic velocity is
fitted to $v_{sys}(LSR)$=1573$\pm$3\kms. The velocity and spatial scales used 
in this paper will be relative to the derived v$_{sys}(LSR)$ and (0,0) 
central offset. At the assumed distance of M100
($D \sim$ 17.1\mpc), 1$\arcsec \sim$ 82\pc.

\section{Molecular gas distribution}

\subsection{The CO maps}

We present in Fig.1 the $^{12}$CO(1--0) velocity-channel maps displaying  
line  emission from v=-91kms$^{-1}$ to v=91kms$^{-1}$. 
No significant emission is found outside this velocity range.
The rms noise
in the channel maps (6.5kms$^{-1}$ wide) is 8.5mK in brightness temperature.
The emitting gas is confined within an elongated region extending 
over $\Delta$x$ \times \Delta$y=20\arcsec $\times$ 30\arcsec.  
As expected for a moderately inclined (i=32$^{o}$) rotating disk whose
 major axis is close to North (PA=153$^{o}$) and assuming that M100's spiral structure
is trailing, gas emission is bound to appear 
mostly at positive velocities on the southern side, contrary to blueshifted 
gas emission showing up on the northern side. Notwithstanding this general trend,
 the morphology of the 
channel maps indicate strong deviations from axisymmetry both in the velocity 
field and in the gas distribution. 
Channel maps with velocities close to v$_{sys}$=0, show a distorted 
S-like pattern reminding of spiral-like streaming.  
Emission from a strong CO source is detected 
within a large velocity range v=(-70 up to 70kms$^{-1}$) at the 
center of the galaxy. There are also evidences of gas emission at 
highly forbidden velocities: CO emission is detected on the 
northern part, towards ($\Delta \alpha$, $\Delta \delta$)=
(+7$\arcsec$,+10$\arcsec$), at positive velocities (v=60 up to 
80kms$^{-1}$).  There is a southern counterpart of this 
component towards ($\Delta \alpha$, $\Delta \delta$)=
(-7$\arcsec$,-10$\arcsec$), at negative velocities (v=-60 up to -80kms$^{-1}$).

\begin{figure}[btph]
\psfig{figure=7230.f1,width=8.5cm,bbllx=0mm,bblly=3mm,bburx=167mm,bbury=220mm,angle=-90}
\caption {$^{12}$CO(1--0) velocity-channel maps observed with the Plateau de 
Bure array with a resolution (HPBW) of 2.2\arcsec x 1.3\arcsec. The dynamical
 center in (1950) equatorial coordinates is indicated by a cross
at $\alpha$=\hmsd 12h20m22.96s, $\delta$=\dms 16d05m57.5s. We display the 
emission in 6.5kms$^{-1}$--wide velocity channels from v=-91kms$^{-1}$ 
up to v=91kms$^{-1}$, symmetrically disposed with respect to the
systemic velocity v=0 (v$_{sys}$=1573kms$^{-1}$(LSR)). Contour levels are 
-20, 20 to 180\,mJy/beam by steps of 20\,mJy/beam.}
      
\end{figure}

\begin{figure}[btph]
\psfig{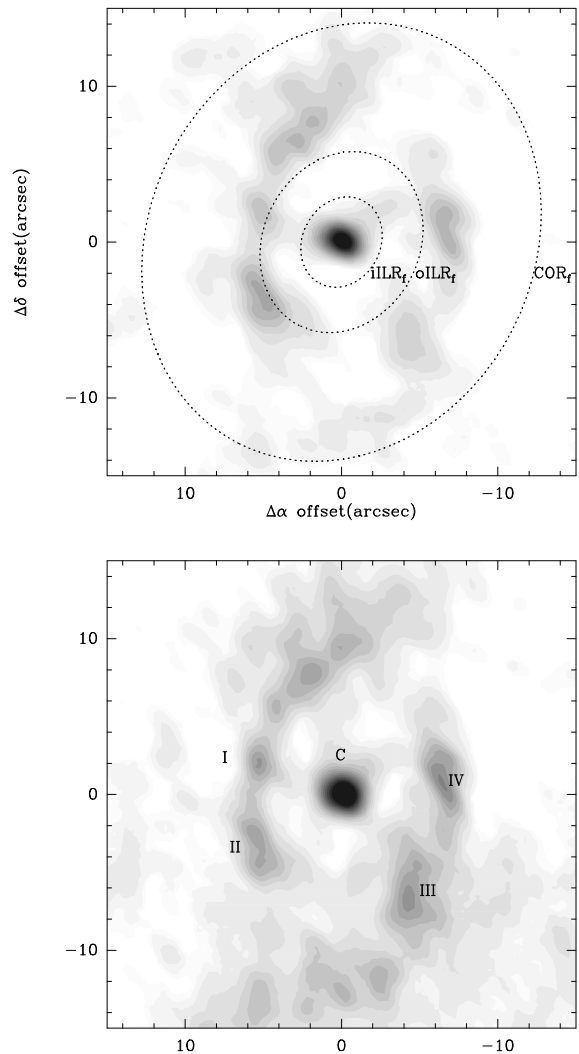}
\caption{ {\bf a (top)} $^{12}$CO(1--0) velocity-integrated intensity contours observed within
the central HPBW=43\arcsec interferometer field of M100. $\Delta \alpha$ and 
$\Delta \delta$ offsets (in arcsec) are
relative to the dynamical center. Gray level contours correspond to 
0.25, 0.5 and 1\,Jykms$^{-1}$beam$^{-1}$ to 10.5\,Jykms$^{-1}$beam$^{-1}$ by steps of
0.75\,Jykms$^{-1}$beam$^{-1}$. Ellipses trace the loci of resonances in the disk (see text) 
{\bf b (bottom)} Same as {\bf a}, but for the combined 
Plateau de Bure+30m map. We indicate the positions of clumps I-IV and C (see text).}     
\end{figure}

The CO velocity-integrated maps (not corrected for primary beam 
attenuation) for the Plateau de Bure (I$_{CO}^{int}$) and the combined 
Plateau de Bure+30m data 
(I$_{CO}^{comb}$) are shown in 
Figs2a-b. They have been obtained by integrating the emission in channels
with brightness temperature higher than 2$\sigma$. The morphology of molecular
 gas distribution seen by the interferometer is characterized by:
 
\begin{itemize}

\item
 
Two spiral arms stretching out from radius r$\sim$7\arcsec and 
position angles PA=130\degr (on the eastern side, hereafter Arm E) and 
310\degr (on the western side, hereafter Arm W). The arms spread over
an azimuthal range of $\Delta \Phi$=120\degr and reach the edges of the map.
Arm W has a sudden change in its pitch angle near PA=210\degr, appearing
broken and discontinuous. 
CO emission along the spiral arms is uneven: it displays a clumpy hierarchy.
Some of these clumps (denoted as {\bf I-IV}) have been spatially 
resolved by the interferometer beam. Spiral arms are narrow: at places, 
they are not resolved transversely (FWHM$<2.2\arcsec$). 
The sign of the observed non-circular motions both in the 
interarm and in the arm regions correspond to the expected velocity pattern 
inside corotation (see discussion of {\bf S95} and section 6).     

\item

There is a strong concentration of molecular gas at the center in the shape of
 an ellipsoidal source which is marginally resolved by the interferometer beam
(major axes--(a,b)=(2.5\arcsec, 2.1\arcsec)). The central clump {\bf C} is connected to the 
western spiral by a bridge of emission.
   
\item

Away from the spiral-like and central sources, gas emission is hardly detected
in the inner and outer {\it interarm} region. 

\end{itemize}

Fig 2b shows the I$_{CO}^{comb}$ map. Roughly the morphology of both 
maps are alike. The zero-spacing flux recovered (defined as 
(f=I$_{CO}^{comb}$-I$_{CO}^{int}$)/I$_{CO}^{comb}$) reaches on average $\sim$50$\%$. 
However, this percentage varies markedly in the map: it lowers up to 
10-20$\%$ for Arm E, it rises up to 50$\%$ for Arm W, and it reaches 
100$\%$ for the interarm region. The flux recovered in the central source 
(f$\sim$30$\%$) makes it appear rounder
 and spatially resolved. Spiral arms are also broader and there is smooth 
interarm emission detected up to r=10\arcsec that the interferometer map 
filtered out completely. Still, the {\bf ND} arm-interarm contrast is high 
($\sim$4--5).

\subsection{Mass of molecular gas}

We derive masses and column densities of molecular gas from I$_{CO}$ using 
a CO-to-H$_2$ conversion factor 
X=N(H$_2$)/I$_{CO}$=2.3x10$^{20}$cm$^{-2}$K$^{-1}$km$^{-1}$s  
(Strong et al. 1988). Assuming the distance to be D=17.1Mpc, the total molecular 
gas mass in the interferometer and combined integrated intensity maps are 
1.4$\times$10$^{9}$M$_{\odot}$ and 2.7$\times$10$^{9}$M$_{\odot}$ respectively. 
The masses of clumps (I, II, III, IV and C; all of them 
of sizes spatially resolved by the interferometer beam) are estimated, first, from 
I$_{CO}^{int}$ and I$_{CO}^{comb}$ (CO based masses including the 
correction factor for helium, M$_{CO}$) and, secondly, from the virial 
theorem (M$_{VIR}$). In the latter case, masses come from 
M(H$_2$)=550$d\sigma_{1d}^{2}$, where d is the equivalent FWHM diameter, 
defined as d=$(d_{1}d_{2})^{1/2}$ ($d_{1}$ and $d_{2}$ are the FWHM 
diameters along the principal axes in a 2D-gaussian model), and 
$\sigma_{1d}$ is the measured radial velocity dispersion seen in the CO line. 
Clumps I, II, III and IV seem gravitationally bound entities 
(M$_{CO}$=1.7-2.7M$_{VIR}$, according to mass estimates from I$_{CO}^{comb}$), 
contrary to the central clump (C), as expected (for C the {\it measured}
 $\sigma_{1d}$ is entirely due to the large velocity gradient of rotation curve). 
CO-based masses range from 0.6 to 2.4x10$^{8}\Msun$ typical of GMA-like 
associations. Note however that we have no accurate calibration of the 
conversion factor in M100 and that the previous conclusions on gravitational 
boundness of GMAs lean on the assumed value of X.

The I$_{CO}^{int}$ contours at half power along the spiral arms correspond on average 
to molecular gas column densities of 400-500$\Msun$/pc$^{2}$. N(H$_2$+He) peaks at 
800-1000$\Msun$/pc$^{2}$ along both spiral arms and it reaches 2000$\Msun$/pc$^{2}$ towards C. 
The critical column densities for gravitational instabilities (N$_c$; Toomre (1964), see also 
Kennicutt (1989) can be derived 
from (see {\bf paper I})

\vspace{2mm}

$N_{c}=\alpha \sigma_v \kappa/G$   

\vspace{2mm}
   
$N_{c}$ range from 100 (at r$\sim$15\arcsec) to 350$\Msun$/pc$^{2}$ (towards C), and 
henceforth the condition for the onset of gravitational instability 
(N(H$_2$+He)/N$_c>$1) is largely fulfilled along the spiral arms and the nuclear source
(N(H$_2$+He)/N$_c\sim$2--6)

We used the relation between I$_{12CO(1-0)}$ and A$_v$ obtained by Cernicharo 
and Gu\'elin (1987) for the Taurus cloud complex to estimate the visual 
extinction towards the center of M100:

\vspace{2mm}

I($^{12}$CO(1--0))=(5.0$\pm$0.5)(2A$_v$-0.5$\pm$0.2) {\bf [1]}  

\vspace{2mm}

where the factor of 2 corrects for the non-obscuring dust behind the HII regions, 
assumed to be located in the middle of the disk. A$_v$ derived from {\bf [1]} ranges 
from 2--2.5 (in the gaseous spiral arms) to 5 (towards C) in good agreement with the average 
value obtained from I($^{13}$CO(1--0)) in {\bf paper I} for the {\bf ND} (A$_v\sim$ 3.5).

\section{Comparison with other tracers}

We discuss in this section the global picture emerging from the comparison of the different
gaseous and stellar tracers in the center of M100. 

The near infrared K pictures of the M100 disk show the presence of an outer stellar bar with
a 90-100\arcsec diameter. A gaseous molecular bar, aligned with the stellar bar,
 has been detected ({\bf paper I}) and 
a flow of gas parallel to the bar is evident in the observed CO
kinematics up to r$\sim$60\arcsec: the CO p-v minor axis diagram shows the 
S-like pattern 
typical of a bar driven gas flow ({\bf GB94}). The molecular gas bar  
has two ridges offset from the bar major axis, 
mimicking the offset leading dust-lanes often observed in barred galaxies.
This indicates the presence of two ILRs, and the tendency of the gas
flow to become perpendicular to the bar in between (Athanassoula 1992).


The previous infrared (I, J and R) images of the nuclear region showed the existence of a 
nuclear oval or inner bar, but with a 30\deg leading angle respect to the outer bar. 
The discrepancy is explained by differential extinction, affecting less severely the K band image. 
The spiral-like distribution of molecular gas and henceforth the extinction maxima locus explain why 
Pierce (1986) see a misaligned nuclear lens. {\bf K95} used the alignment of both K bars as an 
argument supporting the existence of one mode for the disk, opposed to the two independent bars 
scenario. Still, this is just a statistical argument: whatever relative orientation of
two independent bars is equally probable, including the alignment case.    
On the other hand, although the K band picture of M100 is less affected by extinction, the existence of 
substructures such as leading arms or relative maxima at the bar ends (used by Knapen and 
collaborators as supporting evidences of the one mode case), are shown to be linked with extinction
(see below).

\subsection{Bias of extinction}

 Fig.3a shows the overlay of I$_{CO}^{int}$ contours with H$_{\alpha}$, in gray scale. 
Although, similarly to CO, H$_{\alpha}$ emission is confined in a ring-like structure 
and a central source, there are systematic shifts between the CO and H$_{\alpha}$
ridges along the arms. Assuming the spiral to be trailing (see section 6), 
H$_{\alpha}$ appears upstream the CO ridge along both arms for PA=0\degr-90\degr (Arm E) and
PA=180\degr-270\degr (Arm W). On the contrary, the H$_{\alpha}$ ridge lies downstream the CO
maximum in the azimuthal ranges PA=90\degr-180\degr (Arm E) and 
PA=270\degr-360\degr (Arm W). Towards C, H$_{\alpha}$ is relatively dimmed. 
Generally speaking, there is apparently a local anticorrelation between molecular gas 
column densities and massive star formation in
the {\bf ND}: CO peaks suspiciously avoid H$_{\alpha}$ maxima.

From the A$_v$ map we can estimate the opacity in the H$_{\alpha}$ band and 
 correct the map of Fig.3a by a factor e$^{\tau_{H_{\alpha}}}$.  
Although some of the relative shifts between CO and H$_{\alpha}$ clumps 
are still present after correction, many of the reported 
displacements disappear or at least they are considerably attenuated. 
 The latter result indicates that a direct 
comparison between CO and extinction uncorrected H$_{\alpha}$ maps may be 
highly misleading, as expected, considering that A$_v$ ranges 
from 2 to 5, as shown above. 

Additional insight into extinction effects is illustrated in Fig.3b, showing the overlay 
between I$_{CO}^{int}$ (gray scale) and radiocontinuum emission at 6cm (S$_{6cm}$) 
(contours). 
M100 possesses an extended source of radio emission
associated with the {\bf ND}. The brightest radio peak lies at 
$\alpha$ (1950)=12$^{h}$20$^{m}$23.46$^{s}$, $\delta$(1950)= +16$\deg$05$\min$56.5$\arcsec$.
A secondary radio maximum coincides with the optical nucleus.
S$_{6cm}$ contains both thermal free-free emission and non-thermal synchrotron emission. In order 
to estimate the contribution of each component we have considered, first, the existent 
measurements at three 
wavelengths (1.2cm, 6cm and 20 cm) and, secondly, a simple hypothesis on 
the spectral indexes 
of thermal and non-thermal emission ($\alpha_{th}$=-0.1 and $\alpha_{nonth}$=-1.0, respectively, 
where $\alpha$ is defined as S$_{\nu}\,\sim\,\nu^{-\alpha}$).         
The resulting best fit indicates that thermal emission from HII regions dominates at 6 cm, 
as S$_{6cm,th} > $60$\%$. 
As shown in Fig.3b, there is an 
excellent spatial correlation between I$_{CO}^{int}$ and S$_{6cm}$. Henceforth, 
no systematic offsets exist between massive star formation, traced by S$_{6cm}$, 
(not affected by extinction) and molecular gas column densities, traced by I$_{CO}^{int}$.

Fig.4 shows the overlay of I$_{CO}^{int}$ contours with the K-band image (in gray scale) 
which represents the bulk of the old stellar component. 
The isophote deviations in the 
inner 6$\arcsec$ were first interpreted by {\bf K95} as the signature of 
leading arms. A similar interpretation is invoked to account for 
 the K peaks visible at the bar end-points 
(denoted as K1-2). A rough inspection of Fig.4 suggests 
that these features might be affected
by extinction. 
The CO spiral arm ridges going across the K nuclear bar, create 
artificially the weak leading arms signature. The same applies to the K1-2 features. 
Following a similar procedure to the one used for the H$_{\alpha}$ map, we have corrected 
to the first order the K band image by a factor e$^{\tau_{K}}$. We estimate $\tau_{K}$
from A$_K$=0.12 A$_v$ (where A$_v$ comes from 
equation {\bf [1]}). In the corrected map, the K1-2 
features disappear somewhat and the leading arms are less visible.

\begin{figure}[btph]
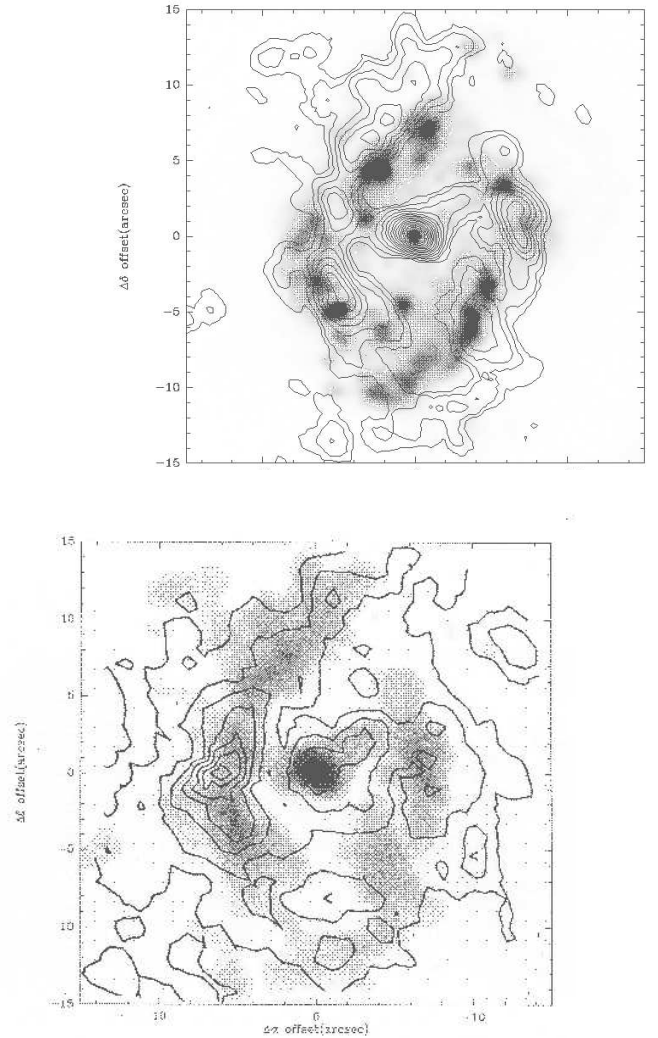

\psfig{figure=7230.f3a,width=8.5cm,bbllx=0mm,bblly=3mm,bburx=190mm,bbury=220mm,angle=-90}
\psfig{figure=7230.f3b,width=8.5cm,bbllx=0mm,bblly=60mm,bburx=190mm,bbury=250mm,angle=90}
\caption {I$_{CO}^{int}$ emission contours (levels as in figure 2) are overlaid  on 
 H$\alpha$ ({\bf a(top)}); {\bf b(bottom)} I$_{CO}^{int}$ gray 
scale contours superposed to 6cm radio continuum emission contours.} 
\end{figure}

\begin{figure}[btph]
\psfig{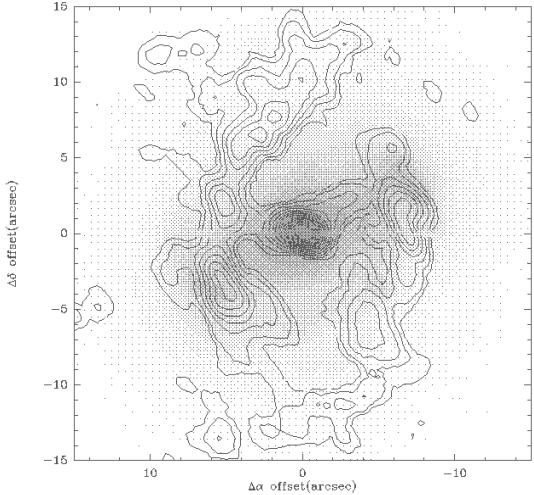}
\caption {We overlay I$_{CO}^{int}$(line contours) on the K-band image (gray scale).}
\end{figure}

\section{Kinematics}

Fig. 5 shows the $^{12}$CO(1--0) isovelocity contours superposed on the I$_{10}$ map. 
Although basically a rotating pattern, the velocity field is characterized by the 
presence of strong streaming motions associated with the nuclear spiral arms. 
Deprojected onto the galaxy plane, the latter reach 
50-60kms$^{-1}$ within the synthesized beam of $\sim 2\arcsec$.

We used the whole data cube to fit both the systemic velocity and the dynamical center, 
as stated in section 2. The global velocity field is highly symmetric with respect 
to v$_{sys}$ and {\bf C}. However, we note that the systemic velocity contour does not go 
exactly across the dynamical center, coincident
with the {\bf C} central clump. As it is seen in the p--v major axis diagram of Fig 6a (see below), 
this is due to the strongly asymmetric wings of the CO emission line towards {\bf C}.

In Fig.6a we show the major axis p--v diagram. 
The deprojected CO rotation curve (v$_{obs}$) derived from Fig.6a 
(assuming i=32$^{o}$) implies a steep rise 
of 180kms$^{-1}$ in less than 2$\arcsec$, associated with the 
ultracompact region {\bf C}, marginally resolved
by the interferometer beam. Henceforth, the velocity gradient there might be higher. The dynamical mass enclosed 
in the inner 100pc is 6.5x10$^{9}\Msun$, the percentage of molecular mass in {\bf C} is $\sim$20$\%$.

At larger radius (at r=$\pm$2--5$\arcsec$) we note the existence of a {\it hole} in the major 
axis p-v diagram, at a position coincident with the interarm crossing, on both sides of the nucleus.   
Therefore we are unable to determine accurately v$_{obs}$ in this region. We can either assume a
 simple solution consisting of
interpolating between the known values of v$_{obs}$ at r=2$\arcsec$ and r=7$\arcsec$, or just consider
v$_{obs}$ to be undetermined within this region. 
The uncertainties of the inner v$_{obs}$ are discussed 
in section 6.2, in particular a comparison with the rotation curve obtained
from the infrared potential(v$_{cir}$) will help to clarify what might be the ``real'' trend of  rotation 
curve in the {\bf ND}. The latter is shown to be crucial in the model fitting.  
Moreover, although tentatively, we report the detection of CO emission from 
two symmetric {\it lobes} at r=$\pm$4$\arcsec$ showing velocities highly forbidden 
by the preponderant rotating pattern (see Fig. 6a).
Isovelocities appear distorted in this region encircling {\bf C} on both sides of the nucleus
(see Fig.5). This could be the signature of orbit crowding in the vicinity of the iILR,
or alternatively out of the plane motions.

Farther away, and coinciding with the crossing of the inner spiral arms (r=10--20$\arcsec$) the 
declivity at both sides of the p--v diagram might evidence strong streaming motions associated with 
the arms.
However the symmetry of the declivity suggest that we see a real decline of v$_{rot}$, 
reflecting the existence of a compact mass distribution in the {\bf ND}. 
Disentangling between the kinematical signatures of the axisymmetric and the 
non-axisymmetric parts of the mass distribution is not straightforward. We illustrate 
below (section 6) how numerical simulations help us in separating 
both contributions. See {\bf paper I} for a discussion of rotation curve at r$>$20$\arcsec$.
     
 We show in Fig 6b the  p--v diagram along the minor axis. We assume the eastern side (x$>$0, 
eastwards) to be the near side. The most remarkable feature is the strength of the radial 
component of the non--circular motions at the crossing of the spiral arms. 
The velocity gradient of the streaming, deprojected onto the galaxy plane, reaches $\sim$80km\,s$^{-1}$ 
within a beam of $\sim$2$\arcsec$ (165pc).

\begin{figure}[btph]
\psfig{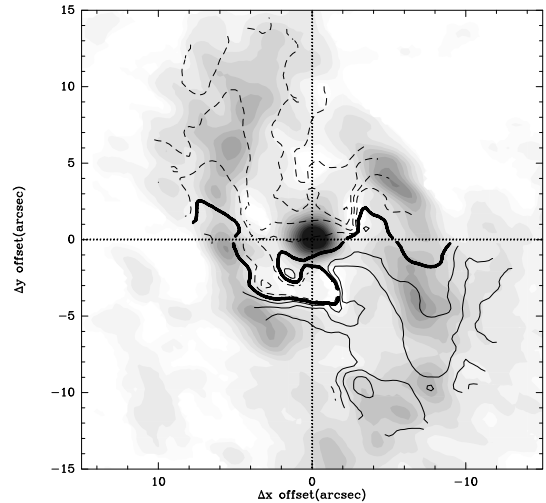}
\caption {overlaid  on a rotated I$_{10}$ grey scale image of M100, oriented 
along the major (y$>$0, northwards) and minor (x$>$0, eastwards) axes, we plot the first-moment 
isovelocity contours derived from the 1--0 synthesized field. Isovelocities go from 
v=-100kms$^{-1}$ to 100kms$^{-1}$, by steps of 20kms$^{-1}$, relative to the
systemic velocity v$_{sys}$=1573kms$^{-1}$(LSR) (displayed by a thick solid 
contour). Dashed contours stand for v$<$0.}
\end{figure}

\begin{figure}[btph]
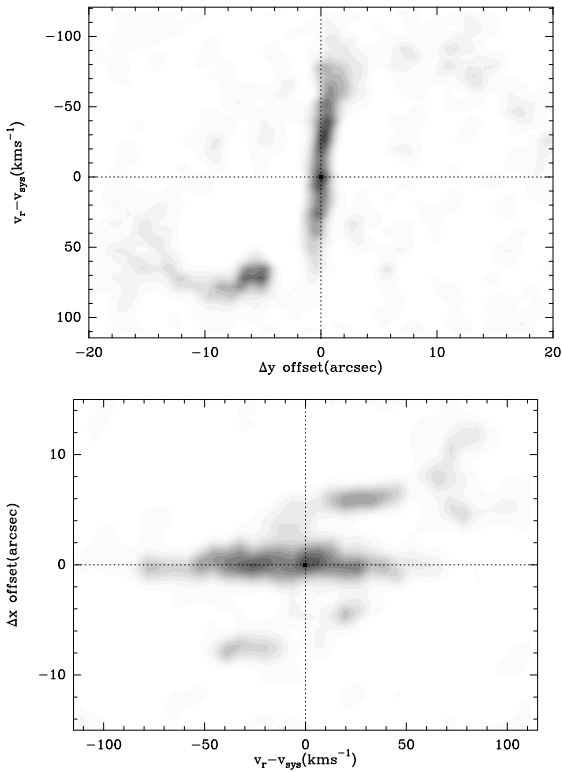

\psfig{figure=7230.f6a,width=8.5cm,bbllx=80mm,bblly=5mm,bburx=190mm,bbury=185mm,angle=-90}
\psfig{figure=7230.f6b,width=8.5cm,bbllx=80mm,bblly=0mm,bburx=190mm,bbury=180mm,angle=-90}
\caption {{\bf a(top)}:The $^{12}$CO(1--0) position-velocity diagram along 
the kinematical major axis oriented with a PA=153$\deg$. 
Y offset (in arcsec) is measured along the major axis. Gray contours are 
linearly scaled from 10$\%$ to 90$\%$ by steps of 10$\%$ of the maximum value.
{\bf b(bottom)}: The equivalent p--v diagram, but now taken along the minor 
axis. X offset (in arcsec) is measured along the minor axis. Same equally 
spaced levels. Dashed lines cross at the dynamical center locus on both figures
(v$_{sys}$, x=0, y=0).} 
\end{figure}

\section{Numerical simulations}

Previous numerical simulations have explored the fit of one pattern associated with 
the primary bar and the spiral structure in the disk of M100 ({\bf G94},{\bf S95}).  
The detailed picture of the nucleus available now at different wavelengths allows to refine
 the model tuning, in particular, to evaluate the feasibility of a two independent patterns explanation
 for the {\it bar within a bar} morphology and to compare the 
different modelling approaches.

We perform simulations
of the molecular clouds hydrodynamics, using a potential as close as 
possible to reality.
This numerical code was developped by Casoli \& Combes (1982) and
 Combes \& Gerin (1985) (see the detailed description of the model in these references) and it was
first applied to a real galaxy (the interacting system M51) by 
Garc\'{\i}a-Burillo et al (1993).

We use galaxy plates of M100 at wavelengths that trace the old population, i.e. 
close to the real mass distribution. The approach is different from what {\bf K95}'s work
is based on for three reasons: first, they use in their simulations 
  a theoretical potential only qualitatively similar to M100's. 
When their theoretical potential and their rotation curve are compared with the directly 
derived from observations the discrepancies are flagrant. Their rotation curve rises much more rapidly
(v$_{rot}$=270kms$^{-1}$ at r=1 kpc), and it reaches a much larger
plateau value farther out (v$_{rot}>$350 kms$^{-1}$ for r$>$ 5kpc).
The position of Lindblad resonances are therefore completely different
in their model and in the real galaxy.
Secondly their simulations use the SPH code, which induces much more
viscosity in the gas behaviour. Henceforth they enhance the action of 
viscous torques compared to gravitational torques. Finally, they intend to fit the gas response 
in the whole galaxy disk by a single fast pattern ($\Omega_p$=70kms$^{-1}$kpc$^{-1}$): 
this is in contradiction with {\bf G94},{\bf S95} who fit successfully  a slow pattern for the outer 
disk. In the present work two models are explored: one slow mode or two coupled modes (slow+fast).

\subsection{Gravitational Potential}

 We have obtained the potential in the plane of the galaxy from the 
combination of a large-scale red image of the outer disk and a small-scale
K-band image of the nuclear region. The red image is 
subject to obscuration in the very dusty
nucleus, while the K-band image is not extended enough to cover
the whole galaxy disk. 
The two images were deprojected onto the plane of the galaxy,
assuming an orientation of PA (position angle)=153$^{\circ}$ and
i (inclination angle)=32$^{\circ}$.
To combine these two images, we performed a Fourier transform at each
radius of the deprojected images, and obtained up to the 8th order
Fourier component (in e$^{i8\theta}$). We plotted the radial
distribution of the coefficients in both images, and found the relative
scale factor to make these distributions continuous. The connection between
the two images was done at 10" ($\approx$800pc). Then the combined image was 
rebuilt from the Fourier coefficients. Higher order coefficients, from $m=4$ 
to 8 were completely negligible inside the nucleus, and their contribution
was slight (a few \% of the $m=2$) but visible, in the large-scale disk.
 
 Through 2D- FFT transform of this recombined image, the gravitational
potential was obtained in the plane. We run the gas simulations in 3D assuming 
cylindrical symmetry for the gravitational forces within the plane (F$_x$, F$_y$).  
This hypothesis is justified considering that the gas thickness (H$_{gas}\sim$50-100pc) 
is much smaller than the stellar thickness (H$_{stars}\sim$1kpc). 
Therefore for the gas itself, we can neglect the variations of both F$_x$ and F$_y$ with z,
making plausible the local approximation of cylindrical symmetry for them.
We assume H$_{stars}$ to 
be constant with radius. The vertical z-forces are derived assuming an isothermal 
stellar disk with a $sech^{2}(z/H)$ density law.
The useful grid used for the potential is 512x512, with a cell of
1" ($\approx$ 100pc), and a maximum radius of $\approx$ 25 kpc. 

\subsection{The rotation curve of the simulations
(v$_{cir}$), compared to observations (v$_{obs}$)}

In a first step, we chose a constant mass-to-light ratio (M/L) to compute the rotation curve
 adopted for the simulations (v$_{cir}$) and
compared with the observed one (v$_{obs}$). The disagreement was severe, especially for the 
inner regions where v$_{cir}>>$v$_{obs}$. The fit can be obtained adjusting M/L as 
a function of radius. It seems logical to think that M/L is to be lowered 
in the {\bf ND} because there might be a non-  contribution of young supergiants
within the K band, which have a low mass-to-light ratio.
Moreover the large amount of dust associated with the high neutral gas column densities
of the nuclear region, which even in the K-band modifies the light profile, also demands 
a modulation of the M/L ratio as a function of radius. 

It is far from straightforward to choose the radial trend of the M/L ratio 
or equivalently the right v$_{cir}$ to introduce in the simulations code.
Imposing v$_{cir}$=v$_{obs}$, seems unrealistic as it implies an anomalous low M/L ratio for the 
{\bf ND} (the implications of this limit case solution are explored in section 6.3.1).
Moreover the determination of v$_{obs}$ is, at places, controversial and it is often
      mixed with strong non-circular motions (see section 5.4).   
 
Still this step is found to be determinant, as the relative positions 
of resonances, and henceforth, 
the morphology of the gas response to the wave, depend strongly, first on the 
pattern speed(s) of the mode(s), and last but not least, on the finally adopted v$_{cir}$. 
Indeed the fit of the optimum rotation curve for the simulations must be envisaged as 
an iterative process, where both Omega$_p$(s) and v$_{cir}$ must be varied {\it coherently} 
to get a sensible fit. In particular when a two patterns scenario is envisaged, we must necessarily 
find an overlap between the resonances of the fast and slow patterns, otherwise the response
 of the gas departs markedly from the observations. 
This coincidence is justified by the theory of modes coupling in 
spiral disks (Tagger et al, 1987; see also Masset and Tagger, 1997). These
restrictions limit considerably the parameter space of the fit when we try 
the solution of two patterns (see section 6.3.2).

Basically we will perform two families of simulation runs: a) with one pattern and   
v$_{cir}\sim$v$_{obs}$ at all radii, and henceforth with a strongly lowered M/L ratio in the 
{\bf ND} and b)with two patterns, dynamically decoupled, for the
inner bar and the primary outer bar. The latter implies a different {\it tuning} of 
the rotation curve in the {\bf ND} (implying v$_{cir}>$v$_{obs}$) and consequently for 
the M/L ratio (see fig.7).

\begin{figure}[btph]

\psfig{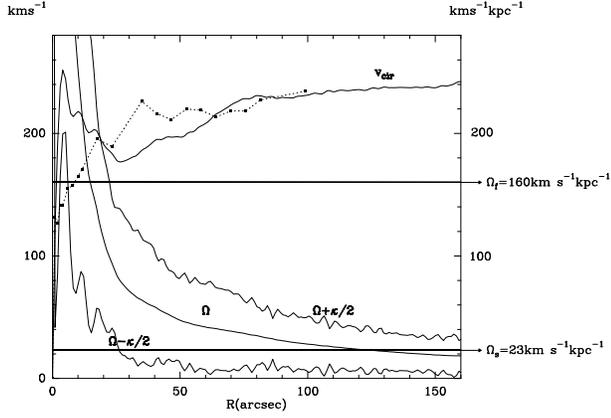}
\caption {We plot the principal resonance frequencies $\Omega$, $\Omega$-$\kappa$/2 
and $\Omega$+$\kappa$/2, derived from $v_{cir}$, i.e., the final rotation curve 
adopted in the simulations. We overlay (with a dashed line) the {\it apparent} rotation curve
obtained from the terminal velocities method (v$_{obs}$). The loci of the resonances are 
determined by the speed of the two patterns: $\Omega_f$=160\,kms$^{-1}$kpc$^{-1}$ 
and $\Omega_s$=23\,kms$^{-1}$kpc$^{-1}$.}
\end{figure}

\begin{figure}[btph]
\psfig{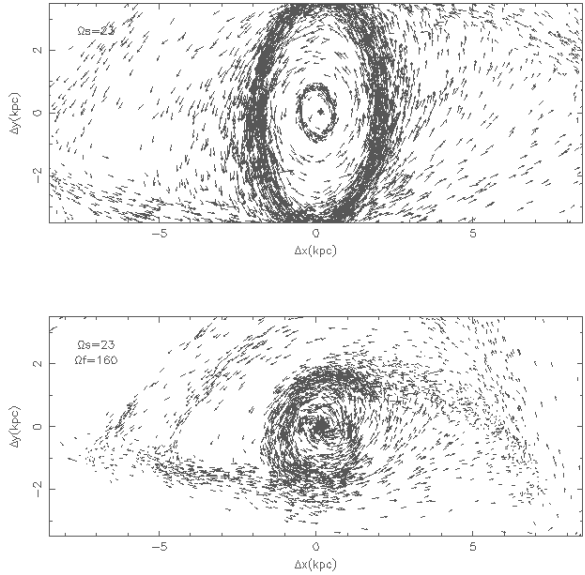}
\caption {We display the particle orbits for molecular clouds in the region where the bar 
instability develops, as they are seen, firstly, {\bf a(top):}
from the frame rotating at $\Omega$=23\,kms$^{-1}$kpc$^{-1}$, for the solution of a single slow pattern
  and {\bf b(bottom):} 
from the frame rotating at $\Omega$=23\,kms$^{-1}$kpc$^{-1}$ for the best-fit solution of a double pattern 
($\Omega_f$=160\,kms$^{-1}$kpc$^{-1}$ (from r=0$\arcsec$ to 10$\arcsec$), 
$\Omega_s$=23\,kms$^{-1}$kpc$^{-1}$(for r$>$10$\arcsec$). The length of the arrows
is proportional to the particle speed in the rotating frame.}  
\end{figure}

\begin{figure}[btph]
\psfig{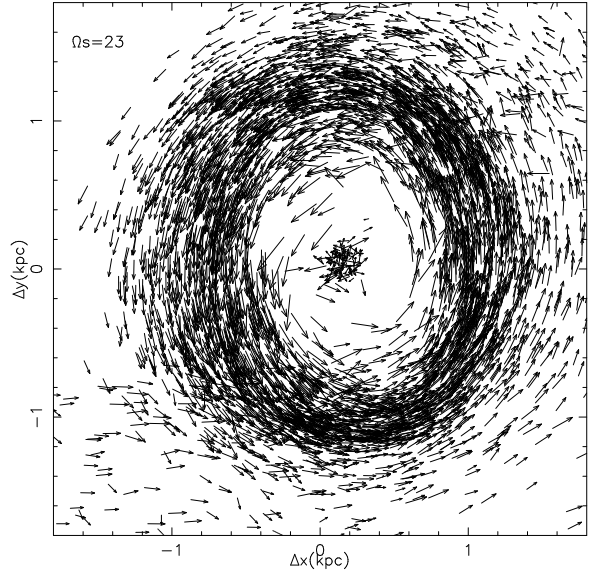}
\caption {We represent the particle orbits in the ND seen from a frame rotating at 
$\Omega$=23\,kms$^{-1}$kpc$^{-1}$, for the slow pattern solution of 6.3.1 but taking v$_{cir}$ tuned for
 the best-fit solution of the double pattern of fig 8b}  
\end{figure}

\begin{figure}[btph]
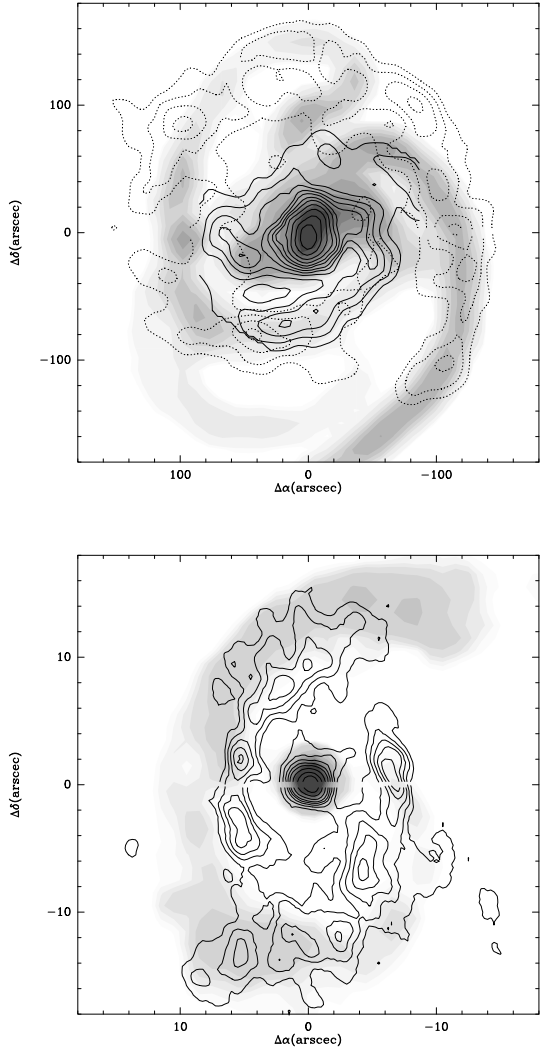

\psfig{figure=7230.f10a,width=8.5cm,bbllx=0mm,bblly=0mm,bburx=190mm,bbury=220mm,angle=-90}
\psfig{figure=7230.f10b,width=8.5cm,bbllx=0mm,bblly=0mm,bburx=190mm,bbury=220mm,angle=-90}
\caption {{\bf a(top):} Superposition of the {\it simulated} gas response (grey scale) with
the molecular gas distribution seen with the 30m (line contours) and the VLA I$_{HI}$ map 
(dashed contours) within the optical disk of the galaxy. {\bf b(bottom):} The same but comparing 
the I$_{CO}$ interferometer levels (line contours) to the simulated gas response (grey scale).}
\end{figure}

\begin{figure}[btph]
\psfig{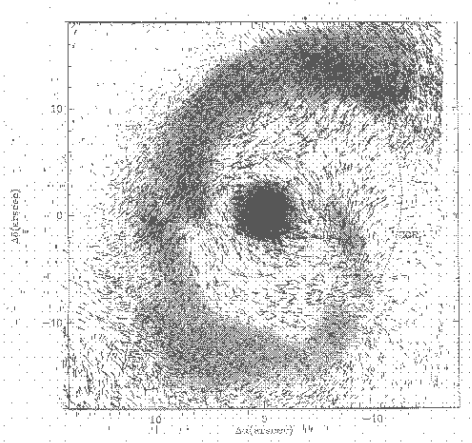}
\caption {Superposition of the particle orbits for the best fit, seen from 
the frame rotating at $\Omega_f$=160\,kms$^{-1}$kpc$^{-1}$ 
(arrows) with the simulated gas response 
(grey scale) in the {\bf ND} of M100. Ellipses trace the loci of the principal 
resonances (COR$_f$, oILR$_f$, iILR$_f$) for the 
{\it fast} pattern, in the epicyclic approximation.}
\end{figure}

\subsection {Best fit solution}
 
We then simulate the hydrodynamics of the gas in the potential 
computed from the combined K and red images. The gas is launched
in equilibrium in circular orbits in the axisymmetrical part of the potential.

 The decomposition of the stellar potential has been done in one
axisymmetric part and two non-axisymmetric parts, to include 
the possibility of two nested bars with two different pattern speeds.
 The nuclear bar in the K-band image has a radius of about 10" ($\approx$ 1 
kpc), and we overlapped the Fourier analysis of the images also at 10".

The axisymmetric part of the density is first used alone to compute
the axisymmetric part of the potential; then we compute the potential
over the whole disk due to the central non-axisymmetric part of the density 
for radii smaller than 10", and a third FFT transform is completed 
for the external non-axisymmetric part of the density ( $r>$ 10").
  The second part of the potential is rotated with $\Omega_f$,
generally at a larger angular speed than the main bar, rotating with $\Omega_s$.
Obviously when  $\Omega_s$=$\Omega_f$, we recover the one mode solution.

  We have not considered here the self-gravity of the gas, since
it only slightly modifies the overall morphology, given the gas to total
mass ratio of NGC 4321 (see {\bf S95}). We have also
run a simple collisional code for the gas, without mass spectrum.
Clouds interact with each other via inelastic collisions,
with their radial relative velocity losing 75\% of their absolute 
value in the collision. 
The collisional grid is two-dimensional, since the thickness of the
gas disk is not larger than the collisional mean free-path.
Clouds are launched with an initial radial distribution which is
exponential with a scale length of 7 kpc (70"). 

  The non-axisymmetric part (one or two bars), rotating as a solid body,
is introduced gradually over a time scale of 150 Myr. Simulations carry on
until a steady response of the gas is reached; on average the condition 
is fulfilled after $\sim$ 900 Myr.

For each numerical run we compute a data cube of the simulated galaxy. We use
 the positions and radial velocities of each cloud at the final time of the 
simulations. The cube is synthesized using a convolution beam similar
to that of the interferometer map, of
FWHM$\sim$2$\arcsec$ for the {\bf ND} and a 30m-like beam of FWHM$\sim$13$\arcsec$ for 
the outer disk.

The gas response is mostly sensitive to the action of gravitational torques
induced by the non-axisymmetric potential of the disk. Depending on the velocity
pattern (s) of the mode (s) and the adopted v$_{cir}$, molecular clouds 
accumulate towards the loci of the principal resonances of the disk, delineating
pseudo-rings, ovals and spiral edges. The morphology of these substructures
and the kinematical signature revealing the population of different orbit families 
during the secular evolution of the gaseous disk, are to be confronted
with the observations and henceforth we derive from this detailed comparison
the optimum fit.

\subsubsection{The one pattern solution}

We extend the slow one pattern solution, found by {\bf G94} and {\bf S95} for the outer bar+spiral 
structure, also to the {\bf ND}, i.e. the two non-axisymmetric parts share the common pattern 
speed: $\Omega_s$=$\Omega_f$=23\,kms$^{-1}$kpc$^{-1}$. 
The main difference introduced by the model developed in this section is the inclusion of a bar component 
in the potential of the {\bf ND}, present in the K band image. Note that {\bf G94} and {\bf S95}
used a red plate where the nuclear bar is not present. Therefore their orbit structure for 
the gas clouds was probably not realistic for the {\bf ND}.

The first family of runs assume a fit on v$_{cir}\sim$v$_{obs}$, following 
the approach of {\bf S95}. In the region where there is no 
information on v$_{obs}$ (between r=2--7$\arcsec$) we interpolate linearly between the 
observed values. In the epicyclic approximation we derive the existence of two ILRs:
a) iILR=0.4 kpc and b) oILR=2.5kpc (see Fig.5 of {\bf S95}).

The response of the gas for the outer disk (r$>$30$\arcsec$) is similar to the best fit solution of 
{\bf S95}: the fit of the outer bar+spiral structure is optimized with 
$\Omega_p$=23 kms$^{-1}$kpc$^{-1}$. In the {\bf ND}, once a 
quasi-stationary state is reached after 800 Myr, an elliptical gaseous ring
appears between r=3.5kpc and 2kpc, i.e, in the vicinity of the oILR.
 An additional and less conspicuous ellipsoidal ring appears between r=0.5 
and 1kpc, i.e., in the vicinity of the iILR.
There is no trace of gaseous orbits aligned along the bar.
The particle orbits, as seen by the rotating frame 
at $\Omega_p$=23 kms$^{-1}$kpc$^{-1}$ are displayed 
in Fig 8a.

The formation of rings near the outer and the inner ILRs are explained by the 
{\it long run} action of gravitational torques, extremely efficient on a 
cloud-structured medium as it is molecular gas. The sign of the torques change at the crossing of 
wave resonances (iILR, oILR, COR and OLR) and also near the maximum of the precession 
frequency $\Omega$-$\kappa$/2. At the latter position gas spiral arms, formed by the relative 
precession of the bar driven orbits, change from trailing to leading. The angular 
momentum transfer to gas leads to the formation of rings near the OLR and also associated with 
the ILRs. A ring can be formed at an intermediate position between the inner and outer ILR or
 eventually two rings associated with each resonance, as seen here.

Although the solution is clearly valid for the outer disk, it fails completely to explain
 the observed morphology of the CO {\bf ND}:
\begin{itemize}

\item

First the size and the orientation of the rings hardly correspond with the observed 
distribution of molecular gas: the gas ring formed near the oILR 
(r$\sim$3 Kpc) is much too 
large compared with the size of the CO {\bf ND} (r$\sim$1-1.5 Kpc).  

\item

We also emphasize the absence of gaseous spiral structure in 
the simulations when they reach steady state. In our model, the gas is stopped at the oILR barrier 
forming a ring-like source at a distance of r$\sim$3 Kpc, and it cannot go inwards. Viscous torques 
acting on the gas at the oILR ring only operate at a much larger time scale and are inefficient to 
drive the gas inwards, opposed to gravitational torques. The morphology is quite similar to the one 
obtained by {\bf K95}   

\item 

The formation of the inner ring at iILR is 
also explained by gravitational torques, still the amount of gas clouds trapped in the inner 200pc
 is clearly insufficient and it cannot account for the strength of the
ultracompact source {\bf C}.       

\end{itemize}

If we adopt v$_{cir}$ directly obtained from a constant M/L ratio (v$_{cir}>>$v$_{obs}$ in the 
{\bf ND}) the oILR locus is at smaller radius (r$\sim$1--1.5Kpc), but although the formed ring 
coincides with the size of the {\bf ND}, no spiral structure is delineated by the gas and no 
ultracompact source is formed at the center (see the {\bf ND} particle orbits in Fig.9). 
With the latter v$_{cir}$ we certainly get the gas 
closer to the center, but not close enough. Moreover the response of the gas departs also from the 
observed one.

\subsubsection{The two patterns solution}

To summarize, the major problem of the one pattern solution is its inability to drive the
 gas to the inner 500 pc and henceforth to account for the observed 
morphology of the CO {\bf ND}. An alternative solution consists of increasing the pattern 
speed of the mode (this is implicitly assumed by {\bf K95} who take $\Omega_p$=70 kms$^{-1}$kpc$^{-1}$). 
That certainly drives the gas inwards, as the principal 
resonances shrink if we take a larger pattern speed. However this solution is incompatible with 
the outer bar+spiral structure fit of {\bf G94} and {\bf S95}): a fast mode solution for the outer disk
 can be excluded. In fact, the isodensity contours seen in Fig 14 of
 {\bf K95}'s model are similar to our fig 8a of our slow one mode solution. The only difference 
is the scale due to their higher pattern speed.

 The high precession rate of the gas orbits in the {\bf ND} strongly suggests the onset of 
a fast mode: the maximum of $\Omega$-$\kappa$/2 reaches 200 kms$^{-1}$kpc$^{-1}$ and a weighted 
average precession rate (derived from 
$<\Omega$-$\kappa$/2$>$=$\int$($\Omega$-$\kappa$/2)I$_{CO}$dr/$\int$I$_{CO}$dr, in the inner  
1kpc) is close to 150$\pm$20kms$^{-1}$kpc$^{-1}$. In the slow pattern solution, the 
rapidly precessing cloud orbits see just an axisymmetric average of the nuclear bar potential, 
and fail to get in phase with the slow bar forcing. The straightforward approach is to try a 
two pattern solution: one slow pattern speed ($\Omega_s$) for the outer disk and a faster mode
($\Omega_f$) for the ND, 
associated with the K-band bar, and whose speed is to be adjusted simultaneously with v$_{cir}$.
Henceforth we build the parameter space of the fit with, first, $\Omega_f$ and secondly,
 with v$_{cir}$ (obtained via a tunable M/L ratio).

We emphasize that the tuning of both cannot be made independently: we explore a range of
pattern speeds between $\Omega_f$=80-180 kms$^{-1}$kpc$^{-1}$, and for a fixed $\Omega_f$
we tune  v$_{cir}$ so as to get spatial overlapping  of corotation of the fast mode
with oILR of the slow mode, otherwise the dynamic decoupling of modes cannot occur
(Tagger 1987; Masset and Tagger, 1997). The choice for the range of $\Omega_f$ (80-180) is not arbitrary: in fact 
outside this range there is no possible overlapping of resonances and no   
 acceptable solutions are to be expected. We underline that the above criterium is a first order 
approach (strictly valid in the epicyclic approximation), and it is taken as a starting point: 
in fact the loci of resonances are 
not circular-like but they can be severely distorted in a barred galaxy({\bf G94}). Also 
they have a broad radial extent and therefore by overlapping we do not mean mathematical 
coincidence (R$_{COR}^{F}$=R$_{oILR}^{S}$).  
In any case, the result of simulations 
will serve as the conclusive test to judge the goodness of the adopted solution.

Among the explored space of parameters the best agreement with the observations is found within 
the interval $\Omega_f$=160$\pm$20 kms$^{-1}$kpc$^{-1}$. 
Outside this range, i.e. below $\Omega_f$=140kms$^{-1}$kpc$^{-1}$, 
 the tuned v$_{cir}$ fulfilling the overlapping condition makes appear anomalous responses in the gas 
for the outer disk, in particular, a strong m=3 pattern outside the large-scale bar. Although {\bf paper I}
shows observational evidences of this pattern (both in the red plate and in the molecular gas response), 
the m=3 mode is weaker than the bar driven m=2 pattern. Moreover the rings or structures formed 
in the gas response depart markedly from the observations and no steady state is reached in the
gaseous disk.      

Fig 8b shows the plot of the particle orbits for the best fit evaluated by a standard 
Chi-squared statistical test: $\Omega_f$=160kms$^{-1}$kpc$^{-1}$. Figs 10a-b display the corresponding
synthesized 30m and interferometer-like cubes. The following observed features 
are accounted for by the best fit solution:

\begin{itemize}

\item
	Formation of a gaseous spiral response in the {\bf ND}  with size and pitch angle
comparable with the observed in the CO mini-spiral. The spiral is formed by precessing x$_1$ orbits 
of the fast pattern, mostly inside its corotation (r=14$\arcsec$, i.e., R$_{cor}^{F}$=1.2kpc). 	
	
\item
	The gas is efficiently driven inwards by the nuclear bar, overcoming the oILR barrier of the slow
 pattern (towards r=25$\arcsec$, leading to $R_{oILR}^{S}$=25$\arcsec$): it falls inwards {\it spiraling} 
from $R_{oILR}^{S}$, crossing 
R$_{cor}^{F}$ and forming an ultracompact source inside the iILR of the fast pattern 
(at r=2.5$\arcsec$, giving $R_{iILR}^{F}$=200pc). The strength and unresolved size of the central source are in excellent 
agreement with the observations.
  
\item

The goodness of the fit for the outer disk (spiral+large scale bar) is maintained.  

\item

Sakamoto et al, 1995 found that the contribution of the nuclear 
bar to the overall potential is dominant in the inner 2 kpc region, and that 
the gaseous spiral structure is mainly driven by the nuclear bar. The rotation 
curve {\it fitted} in our simulations (v$_{cir}$) shows a declining slope
from the center up to r=2 kpc. This is an additional argument supporting the 
idea that the nuclear bar is dynamically decoupled from 
the outer bar and that both entities evolve independently. 
The high precession frequency $\Omega$-$\kappa$/2
in the inner 2 kpc also suggests the onset of a fast mode for the {\bf ND}. 

\item

Fig 11 shows the superposition of particle orbits seen from the fast frame with the 
simulated gas response. Nearly close orbits near R$_{COR}^{F}$ shows the characteristic epicyclic 
shape. Epicycles extend over a broad region (500--800 pc wide in radius), 
and intersect with the ellipsoidal orbits near the oILR of the slow pattern. 
Orbit crowding in the region shared by the two families of  
orbits explain the efficiency of the inwards gas transport.
Moreover, the structure of orbits sustaining the spiral gaseous wave show that there is 
no net upstream or downstream flow of gas all along the spiral arms. This, together with the derived 
high-extinction towards the CO maxima, explains why H$\alpha$ maxima appears either upstream or downstream 
the CO arms showing no regular distribution, contrarily to the predicted gas behaviour in the single 
slow pattern scenario.

\item

There is weak evidence of leading spiral arm structure in our simulations. The onset of a leading arm 
instability appears, as theoretically predicted, 
close to the crossing of the ILR of the fast 
mode (towards r$\sim$2-3$\arcsec$) but it is washed out rapidly 
to form the ultracompact source {\bf C}. Note however that the leading arms would form at a much smaller
 radius than reported by {\bf K95} (they assume a slower pattern 
for the {\bf ND}). 

\end{itemize}    

\section{Summary and Conclusions}

Simulations of the H$_2$ cloud hydrodynamics in the double barred system M100
have shown that the ensemble of observations (optical, infrared, HI and CO maps) are 
best explained by a two {\it independent bars} scenario. The primary stellar bar 
(of 4.5 kpc radius) and the outer spiral structure share a common pattern speed of 
$\Omega_s$=23\,kms$^{-1}$kpc$^{-1}$ which places corotation at R$_{COR}^S$=8-9 kpc, i.e. beyond the 
bar end-points though well inside the optical disk. Although the nuclear stellar bar is 
mostly aligned with the primary bar (within 20\deg) it has been shown to lead a fast pattern 
rotating at $\Omega_f$=160\,kms$^{-1}$kpc$^{-1}$, having corotation at R$_{COR}^{F}$=1.2 kpc 
radius. Both modes are dynamically decoupled and they show overlapping of their major resonances: 
corotation of the fast mode falls well within the ILR region of the slow mode. 

The present model explains the efficient gas transport
towards the nucleus, suggested by the interferometer 
observations, as a consequence of secular evolution driven by the stellar bar. 
Molecular gas crosses the ILR region of the slow pattern, spiraling inwards and forming a 
trailing spiral structure and an ultracompact source encircled by the ILR of the fast 
pattern (R$_{iILR}^{F}$=2.5$\arcsec$). Alternative solutions are unable to account 
for the CO observations. In particular, in the slow pattern solution gas is stopped 
at the ILR barrier and forms a nuclear ring outside the {\bf ND} extent. 
No central gas condensation is formed either.  The fast pattern solution proposed by 
K95 ($\Omega_p$=70\,kms$^{-1}$kpc$^{-1}$) worsens the fit for the outer bar+spiral structure 
found by {\bf GB94}. In addition, two independent methods based on the morphology of the residual 
velocity field for the gas ({\bf S95}) and the identification of spurs in optical pictures 
(e.g. Elmegreen et al 1992) confirm the value of R$_{COR}^S$ reported above.

We conclude that the gas response derived from the CO interferometer map, 
and the relation between the different stellar (K image) and 
gaseous tracers of the {\bf ND} (H$\alpha$) are best explained by the 
two pattern model. In particular, it explains the high CO concentration
in the central part. This gas concentration could be eventually the cause of
the nuclear bar destruction in this fastly evolving galaxy (see Norman et al 1996).

{\it Acknowledgements}.
This work has been partially supported by the Spanish CICYT under grant number 
PB96-0104. We thank J. Knapen for providing us with the H\,I, H$\alpha$ and K-band images
used in this paper.


\begin{thebibliography}{}


\bibitem[1988]{lit2}
Arsenault R., Boulesteix J, Georgelin Y., Roy J.R., 1988, A\&A, 200, 29

\bibitem[1989]{lit2}
Arsenault R., 1989, A\&A, 217, 66

\bibitem[1990]{lit2}
Arsenault R., Roy J.R., Boulesteix J., 1990, A\&A, 234, 23

\bibitem[1992]{lit2}
Athanassoula E., 1992, MNRAS, 259, 345

\bibitem[1992]{lit2}
Canzian B, 1992, PhD, Caltech Astronomy Department (USA) 

\bibitem[1982]{lit2}
Casoli F., Combes F., 1982, A\&A, 110, 287 

\bibitem[1990]{lit2}
Cepa J., Beckman J.E., 1990, A\&AS, 83, 211


\bibitem[1987]{lit1}
Cernicharo J., Gu\'elin M., 1987, A\&A, 176, 299

\bibitem[1985]{lit2}
Combes F., Gerin M., 1985, A\&A, 150, 327

\bibitem[1994]{lit2}
Combes F., 1994. In:  Schlossman I. (ed). Mass Transfer Induced Activity in Galaxies. Lexington
Proceedings.

\bibitem[]{}
Dutrey, A., Ungerechts, H.\ 1995--1996, {\em IRAM PdB Interferometer and 30-m
         Radiotelescope Flux Measurements Reports 11--13}, IRAM document

\bibitem[1992]{lit2}
Elmegreen B.G., Elmegreen D.M., Montenegro L., 1992, ApJS, 79, 37. 


\bibitem[1993]{lit2}
Friedli D., Martinet L., 1993, A\&A, 277, 27

\bibitem[1993]{lit2}
Friedli D., Benz W., 1993, A\&A, 268, 65

\bibitem[1995]{lit2}
Friedli D., Benz W., 1995, A\&A, 301, 649

\bibitem[1994]{lit1}
Garc\'{\i}a-Burillo S., Sempere M.J., Combes F., 1994, A\&A, 287, 419, {\bf GB94}


\bibitem[1993b]{lit1}
Garc\'{\i}a-Burillo S., Combes F., Gerin M., 1993b, A\&A, 274, 148

\bibitem[]{}
Guilloteau, S. and Forveille, T. 1989, {\it Grenoble Image and Line Data
 Analysis System}, IRAM \& Groupe d'Astrophysique, Observatoire de
 Grenoble document

\bibitem[]{}
Kennicutt R.C., 1989, ApJ, 344, 685


\bibitem[1995]{lit1}
Knapen J.H.,Beckman J.E., et al., Heller, C.H., Sholsman I., de Jong R.S., 1995, ApJ, 454, 623 {\bf K95}

\bibitem[1996]{lit1}
Knapen J.H., Beckman J.E., Cepa, J., Nakai N., 1996, A\&A, 308, 27

\bibitem[]{}
Lucas, R.\ 1992, {\it Continuum and Line Interferometer Calibration},
         IRAM document 

\bibitem[1997]{lit1}
Masset F., Tagger M., 1997, A\&A, 322, 442

\bibitem[]{}
Neri, R., Kahane, C., Lucas, R., Bujarrabal, V., Loup, C., 1997, in preparation

\bibitem[]{}
Norman, C.A., Sellwood, J.A., Hasan, H., 1996, ApJ, 462, 114


\bibitem[1986]{lit1}
Pierce J.P., 1986, AJ, 92, 285

\bibitem[1995]{lit1}
Rand R.J., 1995, AJ, 109, 2444

\bibitem[1995]{lit1}
Rauscher B.J., 1995, AJ, 109, 1608

\bibitem[1995]{lit1}
Sakamoto K., Okumura S., Minezaki T., et al., 1995, AJ, 110, 2075

\bibitem[1995]{lit1}
Sempere M.J., Garc\'\i a--Burillo S., Combes F., Knapen J.H., 1995, A\&A, 
296, 45, {\bf S95} 

\bibitem[1997]{lit1}
Sempere M.J., Garc\'\i a--Burillo S., 1997, A\&A, 325, 769 {\bf paper I}

\bibitem[1995]{lit1}
Shaw M., Axon D., Probst R., Gatley I., 1995, MNRAS, 274, 369

\bibitem[1988]{lit1}
Strong A.W., Bloemen J.B.G.M., Dame T.M., et al., 1988, A\&A, 207, 1

\bibitem[1987]{lit1}
Tagger M., Sygnet J.F., Athannassoula E., Pellat R., 1987, ApJL, 318, 43


\bibitem[1964]{lit2}
Toomre A., 1964, ApJ, 139, 1217

\bibitem[1991]{lit1}
Vaucouleurs G. de, Vaucouleurs A. de, Corwin H.G., et al., 1991, 
Third Reference Catalogue of Bright Galaxies. Springer Verlag, New York.


\bibitem[]{}
Vogel, S.N., Wright, M.C.H., Plambeck, R.L., Welch, W.J., 1984, ApJ, 283, 655


\bibitem[1981]{lit1}
Weiler K.W., van der Hulst J.M., Sramek R.A., Panagia N., 1981, ApJ, 243, L151


\bibitem[]{}
Wild, W., 1995, {\em The 30m Manual: A Handbook for the 30m Telescope}, IRAM
         document


\end{thebibliography}
\end{document}